\documentclass[preprint,12pt]{elsarticle}

\usepackage{amssymb}
\usepackage{amsmath}
\usepackage{xcolor}
\usepackage{soul}

\journal{Physics Letters A}

\begin{document}

\begin{frontmatter}



\title{Integrable motion of curves associated with the Fokas-Lenells equation and related spin system} 


\author[aff1]{R. Ramakrishnan\corref{cor1}}
\ead{ramakrishnan.cnld@gmail.com}
\author[aff2]{Sagardeep Talukdar}
\author[aff3]{M. Lakshmanan}
\author[aff4]{Avadh Saxena}

\cortext[cor1]{Corresponding author}


\affiliation[aff1]{organization={Department of Physics, Indian Institute of Technology Kharagpur},
            city={Kharagpur},
            postcode={721302}, 
            state={West Bengal},
            country={India}}
            
\affiliation[aff2]{organization={Department of Physics, Cotton University},
	            city={Guwahati},
	            postcode={781001},
	            state={Assam},
	            country={India}}
\affiliation[aff3]{organization={Department of Nonlinear Dynamics, Bharathidasan University},
	city={Tiruchirappalli},
	postcode={620024},
	state={Tamil Nadu},
	country={India}}
\affiliation[aff4]{organization={Theoretical Division and Center for Nonlinear Studies, Los Alamos National Laboratory},
	city={ Los Alamos},
	postcode={87545},
	state={New Mexico},
	country={USA}}
\begin{abstract}
In this article, we study the gauge equivalence between the integrable Fokas-Lenells equation (FLE) and an associated spin equation through a gauge transformation and the zero curvature condition. We also construct the Lax pair for the generalized spin equation to confirm its integrability. Further, by mapping a generalized spin system on a moving space curve in $\mathbb{R}^{3}$, we show its geometrical equivalence with the FLE. In particular, the associated evolution equations for the curvature and torsion of the space curve are shown to be equivalent to the FLE through a complicated complex transformation unlike the case of the well known Heisenberg spin equation and the nonlinear Schr\"{o}dinger equation.
\end{abstract}



\begin{keyword}


Fokas-Lenells equation, gauge equivalence, integrable spin system, Lax pairs, space curve formalism, integrable motion of curves
\end{keyword}

\end{frontmatter}


	\section{Introduction}

The study of nonlinear integrable systems is a very fascinating area of research due to the mathematical elegance and physical applications. In general, these nonlinear integrable systems possess interesting mathematical structures such as symmetries, conserved quantities, Lax pairs and bi-Hamiltonian structures. In particular the systems admitting soliton solutions, in areas such as nonlinear optics, Bose-Einstein condensation, etc., have attracted  a large number of researchers in these fields over the past few decades.
From the vast sea of integrable nonlinear evolution equations, one of the most well known equations is the nonlinear Schr\"odinger equation (NLSE)
\begin{align}\label{eq:nlse}
	i u_t+\gamma u_{xx}+\rho |u|^2 u=0,
\end{align}
which is used as the standard model for pulse propagation in fiber optics. Here, $\gamma$, and $\rho$ are real constants and $u$ is the complex-valued function of $x$ and $t$. Eq. (\ref{eq:nlse}) is completely integrable and can be linearized by the inverse scattering transform technique\cite{shabat1972exact}. Solutions arising from system (\ref{eq:nlse}) are particularly interesting because of their potential to be used as carriers in long distance communication through optical fibers. However, the NLSE in (\ref{eq:nlse}) can only describe the propagation of pulses up to the first order nonlinear effect, the so called Kerr effect.

To address this issue, more than a decade ago, Fokas and Lenells proposed a more general version of Eq. (\ref{eq:nlse}) when certain higher order nonlinear effects are considered, namely self-steepening, Raman-scattering and third order dispersion. The resultant equation is the following one named as the Fokas-Lenells equation,
\begin{align}\label{eq:fle}
	iu_t-\nu u_{tx}+\gamma u_{xx}+\rho|u|^2(u+i\nu u_x)=0,
\end{align} 
which was deduced in Ref. \cite{Fokas} by means of the bi-Hamiltonian framework. We note here that Eq. (\ref{eq:fle}) and Eq. (\ref{eq:nlse}) are related to each other in the same way that the Camassa-Holm equation is related to the Korteweg de-Vries equation from the bi-Hamiltonian perspective. One can see that Eq. (\ref{eq:fle}) reduces to Eq. (\ref{eq:nlse}) when $\nu=0$. Further, under an appropriate transformation $u\rightarrow e^{ix}u$ and $\nu=\gamma=\sigma=1$, Eq. (\ref{eq:fle}) gets converted to 
\begin{align}\label{eq:fle2}
	u_{tx}+u-2iu_x-u_{xx}-i|u|^2 u_x=0,
\end{align} 
which is integrable in the present form and admits a Lax pair\cite{FL}. The model (\ref{eq:fle}) was derived by A. S. Fokas in Ref. \cite{Fokas},  developed by Lenells in \cite{lenells2010dressing} and its physical foundation was discussed in \cite{Lenells}. An expression for the one-soliton solution of (\ref{eq:fle}) was obtained by the Riemann-Hilbert approach in \cite{FL}. Additionally, through the Hirota bilinear method the authors in \cite{lu2012novel} obtained the soliton solutions, and the rogue wave solution through the Darboux transformation was obtained in Ref. \cite{xu2015n}. \\
\indent It is worth noting that the FLE in Eq. (\ref{eq:fle}) is also equivalent to a very different form under a gauge transformation followed by a change of variables introduced in \cite{Lenells}. Upon the transformation $u\rightarrow\sqrt{\frac{a}{|\rho|}}b e^{i(bx+2abt)}u$, $\xi=2(x+at)$, $\tau=\frac{-ab^2t}{2}$, $a=\gamma/\nu>0$, and $b=1/\nu$, Eq. (\ref{eq:fle}) transforms into another integrable form given as
\begin{align}\label{eq:fle3}
	u_{\tau\xi}=u-2i\sigma|u|^2u_\xi, \quad \quad \sigma=\pm 1.
\end{align}    
From the viewpoint of the spectral problem, Eq. (\ref{eq:fle}) belongs to the derivative NLSE hierarchy that is related to the Kaup-Newell (KN) spectral problem. The form of FLE in (\ref{eq:fle3}) is also well studied in the context of soliton solutions. The $N$-soliton solution through the Hirota bilinear method is investigated in Refs. \cite{Matsuno1,Matsuno2,talukdar2023multi,dutta2023fokas,liu2022fokas,liuoptik} and through the inverse scattering transform method is studied in \cite{lashkincnsns}. Also, the breather and rogue wave solutions of the Eq. (\ref{eq:fle3}) are studied in \cite{zhangsam} using the KP hierarchy reduction method. Its perturbation theory through the inverse scattering technique was discussed in \cite{lashkin2021perturbation}. To avoid any confusion and to analyse the above equations systematically, hereafter in this manuscript we rename Eq. (\ref{eq:fle3}) as type-I FLE and Eq. (\ref{eq:fle2}) as type-II FLE.

The above type of integrable nonlinear evolution equations in (1+1) and (2+1) dimensions have interesting geometrical connections with integrable moving space curves and surfaces \cite{lamb1976, lamb1977}. Such a geometrical connection between the Heisenberg spin system and the NLSE has been effectively demonstrated in \cite{ml1976, ml1977}. Here the geometric relation between the evolution of the curvature and torsion of the moving curve associated with the spin chain mapped onto the NLSE  through a complex transformation, is established. The same geometrical connection has been systematically demonstrated in an entirely different way, namely the gauge equivalence method \cite{tak1977, tak1979}. In this procedure, the Lax pairs of the integrable nonlinear evolution equations and the corresponding integrable spin system are connected through a gauge transformation and the zero curvature equation.

Over the time, these two formalisms have been widely used to understand the integrability property of the nonlinear evolution equations. It is worth noting here that the connection between the Landau-Lifshitz equation and the higher-order nonlinear systems has been studied using gauge equivalence procedure in \cite{anjan1984}. Also, such connections between the next order corrected Heisenberg spin equation and the fourth order Lakshmanan-Porsezian-Daniel equation (LPDE) \cite{lpde1988} as well as the generalized spin equation and the derivative NLSE (DNLSE) \cite{dnlse1987}, have been established in a similar fashion. However, to the best of our knowledge, the geometrical connection for the nonlinear evolution equations having mixed space and time derivatives has not been properly addressed in the scientific literature. To address this issue, in this paper we study the geometrical connection among the type-I and type-II FLEs and the associated spin problems.

This paper is organized as follows. In Sec. 2, we study the gauge equivalence between type-I and type-II FLEs and derive the associated spin systems. As a consequence, from the Lax pairs of the type-I and type-II FLEs, we construct the Lax pairs of the corresponding spin equations, which in turn provide the signature of the integrability of the spin problems. In Sec. 3, we investigate the above mentioned geometrical connection via the so called space curve mapping formalism and suggest the suitable modification in the usual complex transformation is used to exactly map the time evolution of the curvature and torsion equations of space curves to the respective integrable nonlinear evolution equations. Finally, we briefly conclude our main results  in Sec. 4.

\section{Gauge equivalence of the Fokas-Lenells system}
In this section, we introduce the concept of gauge equivalence for both types of FLEs given by Eqs. (\ref{eq:fle2}) and (\ref{eq:fle3}) and show an interesting connection between the considered types of FLE and the generalized Landau-Lifshitz type spin equations. In doing so, we recall that to characterize the integrable evolution of a nonlinear system, one has to typically introduce a pair of associated linear operators, commonly referred to as a Lax pair. The key idea is that the mutual compatibility of these operators governs the dynamics of the nonlinear equation and guarantees its integrability. In particular, the underlying beauty is that upon inserting the Lax pair associated with the nonlinear equation into the compatibility condition also known as the zero curvature equation, it reproduces the original nonlinear system. For instance, we may consider a Lax pair for the system, written in the following compact form:
\begin{equation}\label{linearsystem}
	\Psi_x = U_i\,\Psi, \qquad \Psi_t = V_i\,\Psi,
\end{equation}
where \(U_i\) and \(V_i\) are matrix-valued functions depending on the field \(u(x,t)\), its derivatives, the space--time variables, and a spectral parameter \(\lambda\).  
The compatibility condition \(\Psi_{xt} = \Psi_{tx}\) leads to
\begin{equation}
	U_{it} - V_{ix} + [U_i, V_i] = 0,
	\label{eq:zero-curvature}
\end{equation}
which encodes the nonlinear evolution equation satisfied by \(u(x,t)\). Eq. (\ref{eq:zero-curvature}) is the zero curvature equation as discussed earlier.\\
\indent Let \(G\) denote a Lie group associated with the algebra to which \(U_i\) and \(V_i\) belong.  
A gauge transformation by an element \(g(x,t;\lambda_0)\in G\), defined for a fixed spectral parameter \(\lambda_0\), acts on the Jost function as
\begin{equation}
	\Phi(x,t;\lambda,\lambda_0) = g^{-1}(x,t;\lambda_0)\Psi(x,t;\lambda),
	\label{eq:gauge}
\end{equation}
where the matrix $g(x,t;\lambda_0)$ is,
\begin{align}
	g(x,t,\lambda_0)=\Psi(x,t,\lambda)|_{\lambda=\lambda_0}.
\end{align}
Now, two integrable systems are said to be gauge equivalent if the Lax pair \((U_i,V_i)\) is converted to a new pair \((U_i',V_i')\) through a $\lambda$ independent gauge transformation such as
\begin{equation}
	U_i' = g^{-1}U_i g - g^{-1}g_x, \qquad  
	V_i' = g^{-1}V_i g - g^{-1}g_t,
	\label{eq:gauge-transform}
\end{equation}
and the transformed system satisfies
\begin{equation}
	\Phi_x = U_i'\Phi, \qquad \Phi_t = V_i'\Phi.
\end{equation}
Requiring the consistency of these two relations gives rise to the gauge-transformed compatibility condition
\begin{equation}
	U_i'{_t} - V_i'{_x} + [U_i',V_i'] 
	= g^{-1}\Big( U_{it} - V_{ix} + [U_i,V_i] - (U_{i0t} - V_{i0x} + [U_{i0},V_{i0}]) \Big) g = 0,
	\label{eq:gauge-zero-curvature}
\end{equation}
where \(U_{i0} = U_i|_{\lambda=\lambda_0}\) and \(V_{i0} = V_i|_{\lambda=\lambda_0}\).  
Hence, the new pair \((U_i',V_i')\) generates another nonlinear evolution equation that is gauge equivalent to the original one. Subsequently, to express the transformed nonlinear evolution equation in terms of  
the spin field $``S"$, we introduce,
\begin{equation}
	S = g^{-1} \sigma_3 g; \quad S^2=I,
	\label{eq:spinfield}
\end{equation}
with \(\sigma_3\) denoting the diagonal Pauli matrix.  Then through straightforward differentiation one obtains,
\begin{equation}
	S_x = g^{-1}[\sigma_3, U_0]g, \qquad  
	S_t = g^{-1}[\sigma_3, V_0]g.
	\label{eq:spin-evolution}
\end{equation}
These identities form the basis for constructing explicit Landau--Lifshitz--type spin equations that are dynamically equivalent to the FLE equations.
\subsection{Type-I Fokas-Lenells equation}  
The type-I FLE also known as the Fokas-Lenells derivative nonlinear Schr\"odinger equation (DNLSE) \cite{Matsuno1} is the first negative hierarchy of the DNLSE and is written as 
\begin{align}
	u_{xt}-u+2i|u|^2u_x=0.
	\label{eq:FLE2}
\end{align}
The Lax pair associated with system (\ref{eq:FLE2}) as described by Eq. (\ref{linearsystem}) is
\begin{align}\label{eq:Laxpair3}
	U_1=&\frac{-i\lambda^2}{2}\sigma_3+\lambda u_x,\\ \label{eq:Laxpair4}
	V_1=&\frac{i}{2\lambda^2}\sigma_3 -\frac{i}{\lambda}\sigma_3u+i\sigma_3 u^2.
\end{align}  
The gauge transformation given by Eq. (\ref{eq:gauge}) changes the Lax pair in Eqs. (\ref{eq:Laxpair3})-(\ref{eq:Laxpair4}) to the new form in Eq. (\ref{eq:gauge-transform}) as
\begin{align}\label{eq:u''}
	U_1'=& \frac{i}{2}(\lambda_0^2-\lambda^2)g^{-1}\sigma_3g+(\lambda-\lambda_0)g^{-1}u_xg,\\
	\label{eq:v''}
	V_1'=& \frac{i}{2}(\frac{1}{\lambda^2}-\frac{1}{\lambda_0^2})g^{-1}\sigma_3 g-i(\frac{1}{\lambda}-\frac{1}{\lambda_0})g^{-1}\sigma_3ug.
\end{align} 
Now, after following the steps in accordance with Eqs. (\ref{eq:spinfield})-(\ref{eq:spin-evolution}), we obtain the following relations in terms of the spin field $S = g^{-1} \sigma_3 g, S^2=I$, 
\begin{align}\label{eq:Sx1}
	S_x=&2\lambda_0g^{-1}\sigma_3u_xg,\\
	SS_x=&2\lambda_0g^{-1}u_xg, 
\end{align}  
and,
\begin{align}
	S_t=&-\frac{2i}{\lambda_0}g^{-1}ug,\\
	\label{eq:SSt1}
	SS_t=&-\frac{2i}{\lambda_0}g^{-1}\sigma_3ug. 
\end{align} 
Inserting Eqs. (\ref{eq:Sx1})-(\ref{eq:SSt1}) in Eqs. (\ref{eq:u''})-(\ref{eq:v''}) we finally get,
\begin{align}\label{eq:U''}
	U_1'=& \frac{i}{2}(\lambda_0^2-\lambda^2)S+\frac{1}{2\lambda_0}(\lambda-\lambda_0)SS_x,\\
	\label{eq:V''}
	V_1'=& \frac{i}{2}(\frac{1}{\lambda^2}-\frac{1}{\lambda_0^2})S +\frac{\lambda_0}{2}(\frac{1}{\lambda}-\frac{1}{\lambda_0})SS_t.
\end{align} 
Finally, substituting the above Eqs. (\ref{eq:U''})-(\ref{eq:V''}) into the gauge transformed compatibility Eq. (\ref{eq:gauge-zero-curvature}), we obtain the gauge equivalent form of Eq. (\ref{eq:FLE2}) as
\begin{align}
	S_t-\frac{1}{2i\lambda_0^2}[S,S_{xt}]-\frac{1}{\lambda_0^4}S_x=0, S^{2}=I.
\end{align}
Without any loss of generality, we may choose $\lambda_0 = 1$ and write the equivalent spin equation as
\begin{align}\label{FLE1}
	S_t-\frac{1}{2i}[S,S_{xt}]-S_x=0, S^{2}=I.
\end{align}
Here we note that an attempt has already been made to derive the spin equation corresponding to Eq. (\ref{eq:FLE2}) using the gauge equivalence method in \cite{dutta2023fokas} where the last term was missed. 

\subsection{Type-II Fokas-Lenells equation}
The Type-II FLE being the integrable generalization of the NLSE, preserves the Lax integrability in the following form\cite{FL},
\begin{align}
	u_{xt}+\alpha\beta^2 u-2i\alpha\beta u_x-\alpha u_{xx}-i\alpha\beta^2|u|^2u_x=0.
	\label{eq:FLE1}
\end{align}
The Lax pair \(U_2,V_2\) in accordance with Eq. (\ref{linearsystem}) is,
\begin{align}\label{eq:Laxpair1}
	U_2=&\lambda u_x-i\lambda^2\sigma_3,\\ \label{eq:Laxpair2}
	V_2=& \alpha \lambda u_x+\frac{i\alpha\beta^2}{2\lambda}\sigma_3 u-\frac{i\alpha\beta^2}{2}\sigma_3u^2-i\eta^2\sigma_3,
\end{align}  
where 
\begin{align}
	\eta=\sqrt{\alpha}(\lambda-\frac{\beta}{2\lambda}).
\end{align}
Using the gauge transformation given by Eq. (\ref{eq:gauge}), the Lax pair in Eqs. (\ref{eq:Laxpair1})-(\ref{eq:Laxpair2}) transforms to the new form in Eq. (\ref{eq:gauge-transform}) as
\begin{align}\label{eq:u'}
	U_2'=& (\lambda-\lambda_0)g^{-1}u_xg+i(\lambda_0^2-\lambda^2)g^{-1}\sigma_3g,\\
	\label{eq:v'}
	V_2'=&\alpha(\lambda-\lambda_0)g^{-1}u_x g+i\alpha[(\lambda_0^2-\lambda^2)+\frac{\beta^2}{4}(\frac{1}{\lambda_0^2}-\frac{1}{\lambda^2})]g^{-1}\sigma_3 g+\frac{i\alpha\beta^2}{2}(\frac{1}{\lambda}-\frac{1}{\lambda_0})g^{-1}\sigma_3ug.
\end{align}
Now, following similar steps as above in the case of Eq. (\ref{eq:FLE2}), we obtain the following relations in terms of the spin field $S$ 
\begin{align}\label{eq:Sx}
	S_x=&2\lambda_0g^{-1}\sigma_3u_xg,\\
	SS_x=&2\lambda_0g^{-1}u_xg, 
\end{align}  
and
\begin{align}
	S_t=&2\alpha\lambda_0g^{-1}\sigma_3u_xg+\frac{i\alpha\beta^2}{\lambda_0}g^{-1}ug,\\
	\label{eq:SSt}
	SS_t=&2\alpha\lambda_0g^{-1}u_xg+\frac{i\alpha\beta^2}{\lambda_0}g^{-1}\sigma_3ug. 
\end{align} 
Substituting Eqs. (\ref{eq:Sx})-(\ref{eq:SSt}) into Eqs. (\ref{eq:u'})-(\ref{eq:v'}) we finally get,
\begin{align}\label{eq:U'}
	U_2'=& \frac{(\lambda-\lambda_0)}{2\lambda_0}SS_x+i(\lambda_0^2-\lambda^2)S,\\
	\label{eq:V'}
	V_2'=&(\frac{\alpha \lambda}{2\lambda_0}-\frac{\alpha\lambda_0}{2\lambda})SS_x+i\alpha[(\lambda_0^2-\lambda^2)+\frac{\beta^2}{4}(\frac{1}{\lambda_0^2}-\frac{1}{\lambda^2})]S+(\frac{\lambda_0}{2\lambda}-\frac{1}{2})SS_t.
\end{align} 
After substituting the above Eqs. (\ref{eq:U'})-(\ref{eq:V'}) into the gauge transformed compatibility Eq. (\ref{eq:gauge-zero-curvature}) the gauge equivalent form of Eq. (\ref{eq:FLE1}), which is similar to \cite{myrzakulov2023}, is obtained as
\begin{align}
	S_t+\frac{1}{4i\lambda_0^2}[S,\alpha S_{xx}-S_{xt}]+(\frac{\alpha\beta^2}{4\lambda_0^4}-\alpha)S_x=0, S^{2}=I.
\end{align}
Again, without any loss of generality we may choose $\lambda_0 = 1$, so the gauge equivalent spin equation may be written as
\begin{align}\label{FLE2}
	S_t+\frac{1}{4i}[S,\alpha S_{xx}-S_{xt}]+(\frac{\alpha\beta^2}{4}-\alpha)S_x=0, S^{2}=I.
\end{align}
\section{Space Curve Mapping Formalism}
In this section, we investigate the geometric relationship between the FLE and the generalized spin system using the differential geometric method, namely the space curve mapping formalism. This mapping can be applied between the time evolution of the curvature and torsion of the moving space curve and the spin system using the complex transformation which is also widely known as the Hasimoto transformation for the curve or the Lakshmanan transformation for the spin systems and the method is known as Lakshmanan equivalence \cite{ml1978,ml1979,ml1981,ml2005}.   
\subsection{Type-I Fokas-Lenells equation}
We consider a moving space curve in $\mathbb{R}^{3}$ represented by the orthogonal trihedral $\vec{e_{i}}, i=1,2,3,$ following the Serret-Frenet equations and the time evolution of the trihedral on the space curve
\begin{equation}
	\begin{pmatrix}
		\vec{e_{1}} \\ \vec{e_{1}} \\\vec{e_{1}} 
	\end{pmatrix}_{x}
	=
	L
	\begin{pmatrix}
		\vec{e_{1}} \\ \vec{e_{1}} \\\vec{e_{1}} 
	\end{pmatrix},
	\begin{pmatrix}
		\vec{e_{1}} \\ \vec{e_{1}} \\\vec{e_{1}} 
	\end{pmatrix}_{t}
	=
	M
	\begin{pmatrix}
		\vec{e_{1}} \\ \vec{e_{1}} \\\vec{e_{1}} 
	\end{pmatrix},
	\label{SFE}
\end{equation}
where
\begin{equation*}
	L= 	
	\begin{pmatrix}
		0 & \kappa & 0 \\
		-\kappa & 0 & \tau \\
		0 & -\tau & 0
	\end{pmatrix},
	M=
	\begin{pmatrix}
		0 & \omega_{1} & \omega_{2} \\
		-\omega_{1} & 0 & \omega_{3} \\
		-\omega_{2} & -\omega_{3} & 0
	\end{pmatrix}.
\end{equation*}
Here $\vec{e_{1}}, \vec{e_{2}},$ and $\vec{e_{3}}$ are the unit tangent, normal and binormal vectors. Also, $\kappa=(\vec{e_{1x}}.\vec{e_{1x}})^{2}$ is the curvature and $\tau=\kappa^{-2}\vec{e_{1}}.(\vec{e_{1x}} \times \vec{e_{1xx}})$ is the torsion associated with the curve. 
Now the considered vector form of the spin equation in (\ref{FLE1}) is
\begin{equation}
	\vec{S_{t}}=(\vec{S}\times\vec{S_{xt}})+\vec{S_{x}}.
	\label{fle1}
\end{equation}
Then we associate the spin field $\vec{S}(x,t)$ corresponding to the type-I FLE (\ref{fle1}) with the unit tangent vector $\vec{e_{1}}$, that is $\vec{e_{1}}(x,t)=\vec{S}(x,t)$, and allow it to evolve as (\ref{fle1}) so that we have the evolution equation
\begin{equation}
	\vec{e_{1t}}=(\vec{e_{1}}\times\vec{e_{1xt}})+\vec{e_{1x}}.
	\label{e1t}
\end{equation}
Now using the Serret-Frenet equation (\ref{SFE}), the evolution equation can be rewritten in terms of the trihedral as
\begin{equation}
	\vec{e_{1t}}=-(\omega_{1}\tau+\omega_{2x}-\kappa)\vec{e_{2}}+(\omega_{1x}-\omega_{2}\tau)\vec{e_{3}},
\end{equation}
where $\omega_{1}$, $\omega_{2}$ and $\omega_{3}$ are given below.
Unlike the case of other equations, a complication arises in (\ref{e1t}) as the evolution equation of $\vec{e_{1t}}$ involves the mixed space and time derivatives. But since the ($\vec{e_{1}}, \vec{e_{2}}, \vec{e_{3}}$) form an orthogonal trihedral there is a close relationship. By taking the compatibility condition $L_{t}-M_{x}+[L,M]=0$ for Eqs. (\ref{SFE}), one obtains the following system of coupled nonlinear evolution equations for the curvature and torsion:
\begin{eqnarray*}
	\kappa_{t}&=&\omega_{1x}-\omega_{2}\tau,\\
	\tau_{t}&=&\omega_{2}\kappa+\omega_{3x},\\
	\omega_{2x}&=&-\omega_{1}\tau+\omega_{3}\kappa.
\end{eqnarray*}
In the above equations,
\begin{eqnarray*}
	\omega_{1}&=&\frac{(\kappa-\kappa_{xt})}{(1+\tau)},\\
	\omega_{2}&=&\kappa_{t},\\
	\omega_{3}&=&\frac{\tau(\kappa-\kappa_{xt})}{\kappa(1+\tau)}+\frac{\kappa_{xt}}{\kappa}.
\end{eqnarray*}
The resultant time evolution equations of the curvature and torsion take the following forms
\begin{eqnarray}
	\kappa_{t}&=&-\frac{1}{(1+\tau)}\bigg[\frac{(\kappa_{xt}-\kappa)}{(1+\tau)}\bigg]_{x}, \label{set1.1}\\
	\tau_{t}&=&\bigg[\frac{(\kappa_{xt}-\kappa)}{\kappa(1+\tau)}\bigg]_{x}-\frac{\kappa}{(1+\tau)}\bigg[\frac{(\kappa_{xt}-\kappa)}{(1+\tau)}\bigg]_{x}.
	\label{set1.2}
\end{eqnarray}
We note here that for the well known Heisenberg spin chain equation
\begin{equation}
	\vec{S_{t}}=(\vec{S}\times\vec{S_{xx}}),
\end{equation}  
the associated evolution equations for curvature and torsion, as given in \cite{ml1977}, are
\begin{eqnarray}
	\kappa_{t}=&(\kappa\tau)_{x}+\kappa_{x}\tau,\label{hse1}\\ \tau_{t}=&\kappa\kappa_{x}+\bigg(\frac{\kappa_{xx}}{\kappa}-\tau^{2}\bigg)_{x}.\label{hse2}
\end{eqnarray}
The above coupled nonlinear evolution equations (\ref{hse1})-(\ref{hse2}) can be combined through the complex transformation
\begin{equation}\label{trans}
	q(x,t)=\kappa(x,t) e^{i\int \tau(x,t) dx},
\end{equation}
to obtain the well known nonlinear Schr\"{o}dinger equation,
\begin{equation}
	iq_{t}+q_{xx}+\frac{1}{2}|q|^{2}q=0.
\end{equation}
It is important to note here that the same transformation can not be used for the nonlinear evolution equations which have mixed space and time derivatives like FLE.
Instead the following evolution equations
\begin{eqnarray}
	\kappa_{t}=-\frac{1}{(1+\tau)}\bigg[\bigg(\frac{\kappa(\kappa_{xt}-\kappa)}{(1+\tau)}\bigg)_{x}-4\kappa^{2}\kappa_{x}\bigg], \label{set2.1}\\
	\tau_{t}=\bigg(\frac{\kappa_{xt}-\kappa}{\kappa(1+\tau)}\bigg)_{x}-4\kappa\kappa_{x},
	\label{set2.2}
\end{eqnarray}
are obtained from the type-I FLE (\ref{eq:FLE2}) using the complex transformation $u(x,t)=\kappa(x,t) e^{i\int (\tau(x,t)+1) dx}$. It is evident that there are differences between (\ref{set1.1})-(\ref{set1.2}) and (\ref{set2.1})-(\ref{set2.2}). The associated differences can be eliminated with further careful modification in the complex transformation as $u(x,t)=\kappa(x,t) e^{i\int (\tau(x,t)+b(x,t)+1) dx} e^{i\int c(x,t) dt}$ or even more general transformation of the form $u(x,t)=f(\kappa(x,t),\tau(x,t))e^{i\int \tau(x,t) dx}e^{ia(x,t)}$ and with suitable selection of functions $b(x,t)$ and $c(x,t)$ or $f(\kappa, \tau)$ and $a(x,t)$.\\
\subsection{Type-II Fokas-Lenells equation}
To perform the space curve mapping for the spin equation (\ref{FLE2}), along with $\alpha=\beta=1$ without any loss of generality, corresponding to type-II FLE, we now consider the vector spin equation
\begin{equation}\label{vecspin2}
	\vec{S_{t}}=-\delta(\vec{S}\times\vec{S_{xx}})+\delta(\vec{S}\times\vec{S_{xt}})-(\delta^{2}-1)\vec{S_{x}},
\end{equation}
where $\delta=1/2$. This spin vector $\vec{S}$ can again be mapped onto the unit tangent vector $\vec{e_{1}}$ of the orthogonal trihedral, and so we have the following equation
\begin{equation}
	\vec{e_{1t}}=-\delta(\vec{e_{1}}\times\vec{e_{1xx}})+\delta(\vec{e_{1}}\times\vec{e_{1xt}})-(\delta^{2}-1)\vec{e_{1x}}.
\end{equation}
The above equation can be rewritten, using the standard spatial and time evolution of the orthogonal trihedral (\ref{SFE}) and the relationship among the unit vectors, as
\begin{equation}
	\vec{e_{1t}}=[-\delta(\omega_{2x}+\omega_{1}\tau)+\delta\kappa\tau+(1- \delta^{2})\kappa]\vec{e_{2}}+[\delta(\omega_{1x}-\omega_{2}\tau)-\delta\kappa_x]\vec{e_{3}}.
\end{equation}
As followed in the previous section, the compatibility condition $L_t-M_x+[L,M]=0$ leads to the following system of coupled nonlinear evolution equations for $\kappa(x,t)$ and $\tau(x,t)$ as
\begin{eqnarray*}
	\kappa_{t}&=&\omega_{1x}-\omega_{2}\tau,\\
	\tau_{t}&=&\omega_{2}\kappa+\omega_{3x},\\
	\omega_{2x}&=&-\omega_{1}\tau+\omega_{3}\kappa,
\end{eqnarray*}
where
\begin{eqnarray*}
	\omega_{1}&=&\frac{1}{(1+\delta\tau)}[(1-\delta^2)\kappa-\delta^2(\kappa_{xt}-\kappa_{xx})+\delta\kappa\tau],\\
	\omega_{2}&=&\delta(\kappa_{t}-\kappa_{x}),\\
	\omega_{3}&=&\frac{\delta(\kappa_{xt}-\kappa_{xx})}{\kappa}+\frac{\tau}{\kappa(1+\delta\tau)}[(1-\delta^2)\kappa-\delta^2(\kappa_{xt}-\kappa_{xx})+\delta\kappa\tau].
\end{eqnarray*}
After substituting the value of $\delta=1/2$, we have the evolution equations for the curvature and torsion as
\begin{eqnarray}
	\kappa_{t}=&\frac{1}{(2+\tau)}\bigg[\bigg(\frac{(\kappa(3+2\tau)-\kappa_{xt}+\kappa_{xx})}{(2+\tau)^2}\bigg)_x+\kappa_{x}\tau\bigg], \label{set3.1}\\
	\tau_{t}=&\bigg(\frac{\kappa_{xt}-\kappa_{xx}}{\kappa(2+\tau)}\bigg)_{x}-\frac{1}{2}\kappa\kappa_{x}+\frac{\kappa}{2(2+\tau)}\bigg[\bigg(\frac{(\kappa(3+2\tau)-\kappa_{xt}+\kappa_{xx})}{(2+\tau)^2}\bigg)_x+\kappa_{x}\tau\bigg], \label{set3.2}
\end{eqnarray}
The above two equations of $\kappa_{t}$ and $\tau_{t}$ can not be mapped to Eq. (\ref{eq:FLE1}) through the complex transformation $u(x,t)=\kappa(x,t) e^{i\int \tau(x,t) dx}$ due to the same reason as discussed in the previous section. But consider the following equations
\begin{eqnarray}
	\kappa_{t}&=&-\frac{1}{1+\tau}\bigg[\bigg(\frac{\kappa}{1+\tau}\bigg(\frac{\kappa_{xt}}{\kappa}-\frac{\kappa_{xx}}{\kappa}+\tau^{2}\bigg)\bigg)_{x}+2\kappa^{2}\kappa_{x}-2\kappa_x\tau+\kappa\tau_x \bigg], \\
	\tau_{t}&=&\bigg[\bigg(\frac{1}{1+\tau}\bigg(\frac{\kappa_{xt}}{\kappa}-\frac{\kappa_{xx}}{\kappa}+\tau^{2}\bigg)\bigg)_{x}+2\kappa\kappa_{x}\bigg]. 
\end{eqnarray}
The above equations can be rewritten as
\begin{eqnarray}
	\kappa_{t}&=&-\frac{1}{1+\tau}\bigg[\bigg(\frac{(\kappa_{xt}-\kappa_{xx}+\kappa\tau^2)}{(1+\tau)}\bigg)_{x}
	+2\kappa^{2}\kappa_{x}-2\kappa_x\tau+\kappa\tau_x \bigg], \label{set4.1}\\
	\tau_{t}&=&\bigg[\bigg(\frac{(\kappa_{xt}-\kappa_{xx}+\kappa\tau^2)}{\kappa(1+\tau)}\bigg)_{x}+2\kappa\kappa_{x}\bigg]. \label{set4.2}
\end{eqnarray}
These equations of $\kappa_{t}$ and $\tau_{t}$, will now be able to integrate with the complex transformation $u(x,t)=\kappa(x,t) e^{i\int (\tau(x,t)+1) dx}$ to get Eq. (\ref{eq:FLE1}) for $\alpha=\beta=1$ without any loss of generality. It is very clear that the above two set of equations for $\kappa_t$ and $\tau_{t}$ are not identical. In order to remove the differences that persist in the evolution equations (\ref{set3.1})-(\ref{set3.2}) and (\ref{set4.1})-(\ref{set4.2}), now we have to suitably modify the complex transformation as we proposed in the previous section  $u(x,t)=\kappa(x,t) e^{i\int (\tau(x,t)+b(x,t)+1) dx} e^{i\int c(x,t) dt}$ or even more general transformation of the form $u(x,t)=f(\kappa(x,t),\tau(x,t))e^{i\int \tau(x,t) dx}e^{ia(x,t)}$ and integrate with very particular $f(\kappa, \tau)$, $a(x,t)$, $b(x,t)$ and $c(x,t)$.
We also wish to point out that the analysis carried out for the vector spin equation associated with type-II FLE (\ref{vecspin2}) in \cite{myrzakulov2023} through space curve formalism using the complex transformation (\ref{trans}) will not be valid for the reasons mentioned above.

\section{Conclusions}
In this article, we have succeeded to deduce the generalized spin equation equivalent to the FLE through the well known gauge equivalence method. The integrability of the identified spin system is also confirmed by the presentation of the associated Lax pair. Also, we have mapped the FLE to the corresponding spin system on a suitable moving space curve. Unlike the case of the other evolution equations, the present case requires the simultaneous consideration of the evolution of all three unit vectors constituting the orthogonal trihedral associated with the space curve. We have also given appropriate modifications in the usual complex transformation, used in the Heisenberg spin hierarchy, for the nonlinear evolution equations which have mixed space and time derivatives. We hope the method indicated in the paper will help to identify new integrable spin systems.

\section*{Acknowledgment}
R.R. would like to acknowledge the financial support in the form of DST-ANRF, Government of India for National Post-doctoral Fellowship (File No. PDF/2023/001115).
S.T. wants to thank DST, Government of India for INSPIRE Fellowship (Award No. DST/INSPIRE Fellowship/2020/IF200278).
M.L. wishes to thank the DST-ANRF, Government of India for the award of a National Science Chair position to him (NSC/2020/000029). 
The work of A.S. at LANL was carried out under the auspices of the U.S. DOE and NNSA under Contract No. DEAC52-06NA25396.

\end{document}